\documentstyle[preprint,aps]{revtex}

\begin{document}
\vskip 0.6cm

\centerline{\bf Diffuse Extragalactic Gamma Rays and Gamma Ray Bursts }

\centerline{R.A. V\'azquez$^{1,2}$}

\centerline{\it $^1$ Departamento de F\'{\i}sica de Part\'{\i}culas,}
\centerline{\it Universidad de Santiago de Compostela}
\centerline{\it 15706, Santiago de Compostela, Spain}

\centerline{\it $^2$ Instituto de F\'{\i}sica Te\'orica,}
\centerline{\it Universidade Estadual Paulista}
\centerline{\it Rua Pamplona, 145}
\centerline{\it 01405-900, S\~ao Paulo, S.P.}
\centerline{\it Brazil}

\begin{abstract}
If gamma ray bursts produce a total energy of 10$^{54}$ ergs and this energy 
is concentrated in the high energy tail of the spectrum $E >$ 1 TeV, then they
may account for the observed diffuse extragalactic gamma ray emission for 
energies $> 100$ MeV. Such an energy could be released if the GRB's are 
produced by the burning of the total mass of a neutron star in a phase 
transition which violates baryon number. 
\end{abstract}

\section{Introduction}

The diffuse extragalactic gamma ray flux has cosmological 
consequences since it gives information on processes produced at cosmological 
distances. This flux has been recently measured by EGRET \cite{EGRET_1} in the
energy region from 30 MeV to 100 GeV. The spectrum is obtained by subtracting 
the modelled galactic background. The derived spectrum is consistent with 
a single power law $\alpha = 2.1 \pm 0.03$ and with an intensity of 
1.45~$\pm$~0.05 $10^{-5}$ cm$^{-2}$ s$^{-1}$ sr$^{-1}$ above 100 MeV.

There exist a large number of models to explain the origin of the extragalactic
diffuse gamma ray emission based on discrete sources and on diffuse origin (see
for instance \cite{Pohl_1} for a review). The most favoured model is based on
gamma ray production on unresolved active galactic nuclei \cite{Bignami_1}.
This hypothesis is favoured by the discovery of EGRET that some AGN's are 
strong $\gamma$ ray emitters \cite{Mukherjee_1}. The average spectral index
of the observed blazars agrees with the observed one for the diffuse emission
\cite{Stecker_1}.

In this note we comment on the possibility of a common origin of gamma 
ray bursts (GRB) and the extragalactic diffuse $\gamma$-ray background. 
If GRB's have a high energy spectrum extending up to several tens of TeV or 
higher and if the energy is dominated by the high energy tail, then it may be 
possible that the diffuse spectrum is produced by GRB's. High energy $\gamma$
rays will pair produce on the infrared (IRB) and cosmic microwave backgrounds
(CMBR) and due to magnetic fields the electron positron pair will be delayed, 
producing a continuous diffuse flux. 

In order to see how this is possible let's consider the energy requirements. 
The extragalactic diffuse gamma ray flux between 100 MeV and 100 GeV is well 
represented by a power law of index $\alpha = 2.1 \pm 0.03 $ 
\cite{EGRET_1,Stecker_1,Pohl_1}. This implies a total emissivity of $\dot Q = 
4 \pi \; \Phi(>100 \; \mbox{MeV}) \; d^{-1} \sim 4 \; 10^{46} $ ergs/Mpc$^3$yr,
where we have used $d= 1$ Gpc and $\Phi(>E_{\mbox{\small min}})$ is the 
integral energy flux measured by EGRET.

On the other hand the energy injected by GRB's per unit volume and unit time
is given by $\dot Q_B = E_T R \sim 3 \; 10^{44} $ ergs/Mpc$^3$ yr, where we 
have used the commonly accepted values $E_T \sim 10^{52} $ ergs, the total 
energy produced by a GRB and $R \sim 3 \; 10^{-8}$ Mp$^{-3}$ yr$^{-1}$, the 
rate of GRB's. We see that GRB's do not have sufficient power to produce the 
observed extragalactic flux. However, if a GRB's has significant emission 
at very high energy $E \gtrsim 1 $ TeV, and if the energy flux is dominated by 
this high energy part of the spectrum then we may have sufficient energy to 
produce the observed extragalactic diffuse emission. This view may be 
supported by the recently claim by the HEGRA group \cite{HEGRA} of a 
correlation observed between the GRB 920925c and an excess observed on the same
time and in the same position at energies greater than 16 TeV. A naive 
extrapolation with constant slope of GRB fluxes to this energy would rend a 
flux completely unobservable for HEGRA. If this association is confirmed this 
would imply that the total energy emitted by a GRB's is in fact much higher 
than currently estimated. We can see from above that the total energy required
to accomplish this is of the order of $E \sim 10^{54}$ ergs. This is an 
enormous amount of energy, of the same order of magnitude as the total energy 
of a solar mass. Most of current GRB models could not account for this energy,
but see below.

As it is well known, a high energy photon does not propagate for a long 
distance due to its interaction with the CMBR 
and the IRB. The problem of propagation of photons in the intergalactic medium 
has been extensively studied \cite{Stecker_2,Coppi_1}. We follow closely the 
conclusions of Aharonian and Coppi \cite{Coppi_1}. Any high energy photon 
injected at cosmological distances will pair produce in the CMBR. For a 1 TeV 
photon the mean free path is small, typically less than a few hundred Mpc. 
Subsequent cascading will produce many photons with energies well below the 
TeV. The important point to notice is that the spectrum of the produced photons
is independent of the original injection spectrum. The general features of the
spectrum are as follows \cite{Coppi_1}: For $E < E_b$ the reprocessed spectrum 
goes like $E^{-1.5}$ and for $E_b < E < E_{\mbox{\small cut}} $ the spectrum 
goes like $E^{-\alpha}$, where $\alpha \sim 2$. $ E_{\mbox{\small cut}} $ is 
the cut off energy due to interaction with the CMBR and IRB, and $E_b \sim 1 
\; \mbox{GeV} \; (E_{\mbox{\small cut}}/1 \; \mbox{TeV})^2 $. Most of the 
energy is concentrated in the second region, where we are interested. 

{From} this we see that the observed extragalactic diffuse gamma ray flux in 
the region from 100 MeV to 100 GeV is consistent with being produced by the 
cascading at cosmological distances of an unknown and otherwise arbitrary flux.
Using $E_b \sim 100 $ MeV, we obtain $E_{\mbox{cut}} \sim 300 $ GeV. Note that 
if we assume that the injection took place at $z \sim 1$ then $E_{\mbox{cut}} \
sim 300 $ GeV is in agreement with the prediction by MacMinn and Primack 
\cite{Primack_1} based on galaxy formation. This result is also consistent with
the upper limit on the IRB from the detection of Mkn 501 by the HEGRA 
collaboration \cite{HEGRA_2}. Previous models would have for $z \sim 1$, 
$E_{\mbox{cut}}$ values as low as 10 GeV (see figure 2 on ref.\cite{Coppi_1}).

The other effect of cascading in the IRB is the delay on the arrival time. This
effect has been already extensively studied \cite{Plaga_1}. In the presence of
an extragalactic magnetic field the pair produced by the original photon is
bent and the increased distance will delay the cascade photons with respect to 
the original burst. Due to fluctuations on the propagation, there 
will be a spread on time of the electromagnetic cascade. The spread time is
of the same order of magnitude as the delay time itself and both are given by 
\cite{Plaga_1}:
\begin{equation}
\Delta t \sim 6.5 \; 10^5 \frac{d}{\mbox{Gpc}} \left( \frac{E_\gamma}
{\mbox{100 MeV}} \right)^{-2} \left( \frac{B}{10^{-18} \; \mbox{G}} \right)^2 
\; \mbox{years},
\end{equation}
where we are normalizing to the minimum energy, $E \sim 100$ MeV. On the other
hand the spread time must be less than the age of the universe, if we want to
observe today the spreaded flux. This gives the condition:
\begin{equation}
B < 10^{-16} \; \mbox{G}.
\end{equation}
Any magnetic field higher than this would invalidate the model since, in its
presence, photons of very low energy would not have had enough time to reach 
us.

Assuming then that the above holds, the spreaded (observed) flux will be given
by:
\begin{equation}
\frac{dN^{obs}}{dE} = \frac{d}{4 \pi} \; E_T \; R \; K \; E^{-\alpha} =
4 \; 10^{-7} \; \left(\frac{E_T}{10^{54} \mbox{erg}} \right) \; \left(
\frac{R \; \mbox{Mpc}^{3} \mbox{yr}}
{3 \; 10^{-8} } \right) \; \left( \frac{d}
{1 \mbox{Gpc}} \right) \; E^{-\alpha}
\; \frac{\mbox{ph.}}{\mbox{cm}^2 \mbox{s sr GeV}};
\label{flux_1}
\end{equation}
where $K$ is a constant which depends on the minimum energy, $d$ is the
average distance of a GRB and $E_T$ is the total energy liberated by the GRB.
For typical values of the parameters the predicted flux agrees well with the 
observed spectrum both in normalization and slope.

Another way to see this is 
the following. The observed diffuse spectrum is given by the sum of all 
the delayed fluxes produced by GRB's in a given time. i.e.:
\begin{equation}
\frac{dN^{obs}}{dE} = \frac{1}{4 \pi} N_{\mbox{\small GRB}} \; \; \frac{\tau}
{\Delta t} \; \; \frac{dN^{eq}}{dE},
\end{equation}
where $N_{\mbox{\small GRB}}$ is the number of GRB's which occur in a time 
$\Delta t$, $\tau$ is the mean duration of a GRB and $\Delta t$ the delay 
(spread) time. $ dN^{eq}/dE$ is the flux of photons produced by the cascading 
of the original flux of a single GRB. The factor $\tau/\Delta t$ takes into 
account the dilution of the flux due to the spread on time. The important 
point to notice is that the observed flux is in fact independent of $\Delta t$
since we are integrating over a total of $N_{\mbox{\small GRB}} \sim \Delta t /
T$ bursts, where $T \sim 1 $ per day is the observed rate of GRB's. This 
estimate is equivalent to the previous one, eq. (\ref{flux_1}). 

Is it possible to have GRB's producing 10$^{54}$ ergs? For most models this 
is a difficult task \cite{Piran_1,Rees_1,Nemiroff_1}. However, this can be 
achieved in scenarios where neutron stars, due to gravitational instability, 
undergo a phase transition \cite{Turok_1}. The result is that a burning front 
is produced which could convert into radiation the whole neutron star. For 
such models we expect a total energy of order $E \sim 5 \;10^{54} 
(M_{\rm tot}/3M_\odot)$ ergs, where $M_{\rm tot}$ is the total neutron mass. 
In addition this model explains easily the baryon load problem since the 
burning front is very efficient in transforming baryons into leptons.

The compactness problem can be resolved in the usual way assuming a 
relativistic shock. The source will be optically thin if it is moving towards
us with a lorentz factor \cite{Piran_1} $\gamma > 10^{18/(4+2\alpha)} \;
(E_T/10^{54} \;  {\rm ergs}) \sim 300 $. Where we are assuming that the 
spectral index, $\alpha \sim 1.5$, in order for the flux to be dominated by 
the high energy tail. This does not modify substantially the constrain on the 
$\gamma$ factor.

In this respect it is interesting to note that, using common values for the 
total energy and GRB rate, $E_T=10^{52}$ ergs and $R= 3\; 10^{-8}$ Mpc$^{-3}$ 
yr$^{-1}$, HEGRA would not be expected to observe any GRB's at TeV energies 
\cite{Mannheim_1}, as the predicted rate for observation of high energy events 
associated to GRB by  HEGRA is 1 every 100 years \cite{Mannheim_1}. Therefore 
the observation of 1 event over a period of 1 year (the analyzed time), 
suggests either a flux or a burst rate 100 times larger, in complete 
accordance with our proposed results. Moreover, if we assume that the event 
GRB 920925c took place at around 100 Mpc (and this is necessary in order for 
HEGRA to detect it), the flux observed by HEGRA is in agreement with an 
extrapolation of typical GRB fluxes at low energy, with a flat spectrum, 
$\alpha \sim 1.5$. This implies again that most of the energy goes into the 
higher energy part of the spectrum. On the other hand, extrapolation of the 
fluxes measured by EGRET to TeV energies with a steeper slope do not give any 
measurable flux at HEGRA. 

This result is further supported by possible detection of other GRB by EAS-TOP
\cite{EASTOP} and by an array at the University of Dublin \cite{Plunkett_1}.
In both cases a detection would be impossible with the currently accepted 
values of total energy and rate of GRB \cite{Mannheim_1}.

There are a number of predictions which make the model testable. If GRB's 
produce the diffuse gamma ray flux then AGN's should produce a negligible 
amount of power in gamma rays, which should be tested in a near future. Also, 
the model of GRB's formation predicts a large flux of high energy ($\sim $ 
100 GeV neutrinos), essentially 10$^{52}$ ergs go to these neutrinos 
\cite{Turok_1}. This flux could be observed, in principle, in large detectors 
such as AMANDA \cite{Halzen_1}. In fact the number of interacting neutrinos in 
the fiducial volume would be large, of order of 100 \cite{Halzen_1}, enabling a
detection with a high statistical significance. Finally, if significant power 
is produced at TeV energies by GRB's we should observe more correlated events
of GRB's with air shower arrays at ground.

\centerline{Acknowledgements}

The author thanks C.O. Escobar, F. Halzen, M.C. Gonzalez Garcia, and E. Zas 
for useful discussions. I thank the IFT for its kind hospitality where this 
work was done. This work was supported by FAPESP.

{\it Note}
When completing this work we were aware of similar work by T. Totani, 
astro-ph/9810206, astro-ph/9810207. He came to similar conclusions as ours 
using different arguments. In particular he relates the UHECR 
producuction to GRB's and arrives to a similar energy requirements. The 
differences are due mainly to a different GRB rate and the unknown distance 
scale.

\end{document}